\documentclass[aps,prb,twocolumn,showpacs,epsfig]{revtex4}
\usepackage{graphicx} 
\begin{document}
\title
{Growth of Tellurium on As-exposed Si(211)}

\author{Bikash C Gupta,  Inder P. Batra and S. Sivananthan}
\affiliation{ Department of Physics, 845 W Taylor street,
University of Illinois at Chicago, Chicago, Illinois 60607-7059,
USA}

\date{\today}

\begin{abstract}
Electronic structure calculations are performed to obtain the
As-exposed Si(211) and the Te adsorbed As-exposed Si(211) surface.
Arsenic-exposed Si(211) may be obtained by adsorbing As on Si(211)
or by replacing surface Si atoms by As. First, we carry out
systematic investigations to obtain stable As-exposed Si(211) due to
As adsorption at various coverages. We find that at 1/2 monolayer
(ML) coverage of As, the highly terraced Si(211) surface becomes
flat decorated with parallel As chains extending along the
[$01\bar{1}$] direction. At 1 ML coverage the Si surface
essentially retains its ideal structure with an added layer of As.
Motivated by the adsorption sequence in the HgCdTe (MCT) growth on
Si, Te adsorption on such an As-exposed Si(211) is studied and 1/2
ML of Te coverage is found to be energetically feasible. Next, we
explore a stable As-exposed Si(211) upon replacement of surface Si
atoms by As. An energetic comparison reveals that the As-exposed
Si(211) obtained by replacing surface Si atoms with As is more
favorable compared to that obtained by adsorbing As on Si(211). In
line with the adsorption sequence in the MCT growth on Si, Te is
then adsorbed on the most favorable As-exposed Si(211) and in
contrast to earlier situation, Te coverage here is found to be 1/4
of ML which agrees with the experiment.
\end{abstract}


\maketitle

\section{Introduction}

As technological devices are often developed on Si substrates, Si
surfaces continue to be a subject of intense theoretical and
experimental studies. Extensive investigations have been done on the
low index Si surfaces, for example, Si(001) and Si(111).
\cite{book1,book2} Most of the theoretical investigations have been
confined to low index surfaces due to the `simplicity' of the
surface. However, the high index surfaces, Si(211), Si(311),
Si(331), Si(557), and Si(553), have attracted some attention
recently \cite{chadi,olshanetsky,nibir0,rujirawat,yang, smith,
thesis,baski0,baski,baski1,baski2,himpsel1,WKI,berghaus,wang,
kaplan,sen}. High index surfaces play technologically important role
as substrates for the fabrication of long wavelength infra-red
detectors \cite{rujirawat,yang, smith, thesis} and as substrates for
the formation of metallic nanowires \cite{baski1,baski2,himpsel1}.

\begin{figure}
  \includegraphics[width=3in]{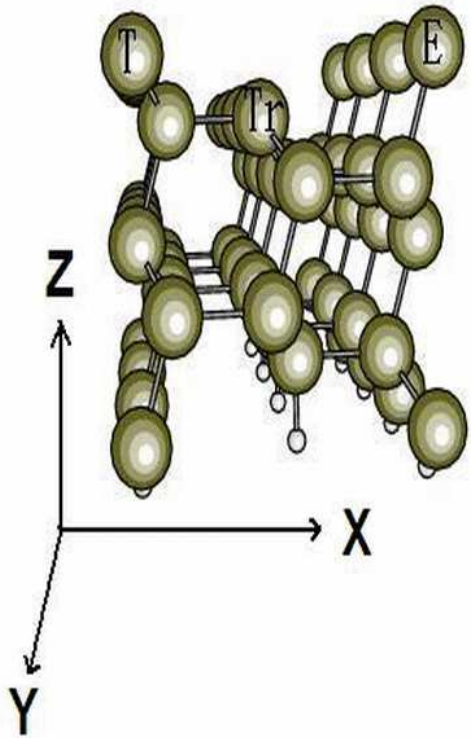}\\
  \caption{Atoms in perspective form the ideal Si(211) super-cell. The
  bottom layer Si atoms are passivated by hydrogen atoms (small circles).
  Surface terrace, trench and edge atoms are denoted as T, Tr
  and E respectively. X, Y and Z directions
  correspond to [$\bar{1} 1 1$], [$0 1 \bar{1}$] and [$2 1 1$]
  respectively }
  \label{fig1}
\end{figure}

Many of the high index Si surfaces are complicated in structure due
to the existence of terraces and steps. The surface of interest here
is Si(211) which can be looked upon as stepped arrangement of narrow
(111) terraces. A three dimensional view of a small portion of the
ideal Si(211) surface is shown in Fig.~\ref{fig1}. The atoms marked
T (called the terrace atoms) on the terrace are three-fold
coordinated and thus have one dangling bond each; those on the step
edge, marked E (called the edge atoms) are two-fold coordinated and
have two dangling bonds each. Atoms in the second layer and closest
to the edge atoms are denoted as Tr (called the trench atoms) have
one dangling bond each. The Si(211) surface consists of two-atom
wide terraces between terrace and edge atoms along the [$\bar{1}11$]
direction. Two consecutive terraces are separated by steps and are
9.4 \AA~ apart in the [$\bar{1}11$] direction, while they extend
infinitely along [$01\bar{1}$].

The Si(211) surface is now one of the surfaces of choice for
epitaxial growth of polar (both III-V and II-VI) semiconductors on
Si. It has been shown earlier~\cite{WKI} that the Si(211) surface
leads to a better quality epitaxial growth of GaP as compared to
Si(001) because it satisfies both the requirements of interface
neutrality and offering inequivalent favorable binding sites for Ga
and P. The Si(211) surface has atoms with both one and two dangling
bonds. The atoms with two dangling bonds can accommodate P, whereas
Ga binds with Si(211) that has a single dangling bond. Large area
high quality CdTe layers have also been grown on the Si(211) surface
for subsequent growth of HgCdTe.~\cite{rujirawat, yang, smith} In
particular, our motivation for studying Si(211) is due to emerging
experimental interest in epitaxial growth of HgCdTe after a
successive growth of ZnTe and CdTe on the As-exposed surfaces aiming
at the development of large area focal plane arrays for the
fabrication of detectors. \cite{thesis}

As far as the reconstruction of Si(211) is concerned there have been
several studies.\cite{berghaus,wang,kaplan,sen} However, a recent
study by Baski {\it et al.} \cite{baski0,baski} was conclusive where
the authors showed in their STM images that the clean Si(211) is
unstable and it consists of nanofacets with (111) and (337)
orientations. As there are evidences \cite{baski1} that the (211)
orientation is regained due to metal adsorption on Si(211), we will
use the bulk terminated surface to study the the As adsorption on
Si(211). In addition, we will extend our study to understand the
atomic configurations of Te on the As-exposed Si(211).

It has been established in recent experiments \cite{thesis,romano}
that the epitaxial growth of II-IV materials on an As terminated Si
surface gives a better quality film compared to that on a bare Si
surface. For example, high quality interface and better ZnS films
were obtained \cite{romano} with As-exposed Si(001). A better
quality MCT growth is possible after a successive growth of ZnTe and
CdTe on the As-exposed Si(111) and Si(211) surfaces.
\cite{thesis,chad} So far, we do not have a comprehensive
understanding of the interaction and atomic configuration of As on
the Si(211) surface. We therefore, carry out extensive electronic
structure calculations for As adsorption on the Si(211) surface at
various coverages. Here we note that an earlier study \cite{sen} to
understand the interaction of As on Si(211) revealed important
results but was incomplete. There are speculations \cite{brill}
based on the analysis by Bringans \cite{bringans} that As atoms may
replace surface Si atoms instead of being adsorbed on Si(211). So,
we perform further calculations by replacing the terrace, trench and
edge Si atoms by As. Comparing all the results and also based on
physical grounds we provide the energetically favorable surface
structures and As positions on the Si(211) surface at various
coverages.

Recently \cite{nibir}, Auger-electron spectroscopy was used to
determine the Te coverage on an As-exposed Si(211) surface and it
was found that only approximately 20-30 \% of a monolayer (ML) of Te
is bonded to the Si(211) surface. This experimental result
encouraged us to further carry out energetic calculations for the Te
adsorption on the As-exposed Si(211). Our total energy calculations
reveal that Te coverage on the As-exposed Si(211) is preferably 25\%
of 1 ML which agrees with the experimental result.\cite{nibir} To
the best of our knowledge, no such calculations exist in the
literature.

The paper is organized as follows. In Section~\ref{sec:method} we
present the parameters used in the pseudopotential density
functional calculations. The results and discussions are presented
in Section~\ref{sec:results}. The results for the As adsorption on
the Si(211) are presented in subsection \ref{subsec:AsAdsorption}.
Subsection~ \ref{subsec:TeOnAsAdsorbedSi} gives the results for the
Te adsorption on the As adsorbed Si(211), \ref{subsec:AsReplacement}
contains the results for Si replacement by As and~
\ref{subsec:TeOnSiRepByAsSi} give results for Te adsorption on
Si(211) surfaces with all the terrace and trench Si atoms being
replaced by As. And finally, in section~ \ref{sec:summary}, we
summarize our principal results.

\section{Method}
\label{sec:method} Total energy minimization calculations are
carried out within the density functional theory (DFT) in
conjunction with the pseudopotential approximation. The Si(211)
surface is represented in a repeated slab geometry. Each slab
contains seven Si(211) layers with a vacuum region of 12 \AA. Each
layer contains 8 Si atoms--2 along [$\bar{1} 1 1$] and 4 along
[$01\bar{1}$]. The top layer contains 4 edge and 4 terrace Si atoms
as shown in Fig.~\ref{fig1}. It is noted that in an ideal Si(211) $1
\times 1$ surface each layer consists of two distinct atoms. The Si
atoms in the bottom layer have their dangling bonds saturated by H
atoms (see Fig.~\ref{fig1}). Since the edge atoms have 2 dangling
bonds each, the trench atoms have 1 dangling bond each and the
terrace atoms have 1 dangling bond each, we require 16 H atoms to
saturate all the dangling bonds at the bottom of the slab. The top
five Si layers are relaxed for geometry optimization while the two
lower-most Si and the H layers are held fixed to simulate the
bulk-like termination. The wave functions are expanded in a plane
wave basis set with a cutoff energy $|\vec{k} + \vec{G}|^2 \le 250$
eV. The Brillouin zone (BZ) integration is performed within a
Monkhorst-Pack (MP) \cite{mankefors} scheme using 4 inequivalent
k-points. It has been established earlier \cite{sen} that this
energy cutoff and k-points give sufficiently converged values for
the binding energies. Ionic potentials are represented by
Vanderbilt-type ultra-soft pseudopotentials~\cite{ultrasoftpp} and
results are obtained using generalized gradient approximation
(GGA)~\cite{pw91} for the exchange-correlation potential.
Preconditioned conjugate gradient is used for wave function
optimization and a conjugate gradient for ionic relaxations. The Z
axis is taken perpendicular to the Si(211) surface, while X and Y
axis are along [$\bar{1} 1 1$] and [$0 1 \bar{1}$] respectively. The
VASP code ~\cite{vasp} is used for our calculations.

\section{Results and Discussions}
\label{sec:results}

The results for the adsorption of As on Si(211) at 1/8, 1/4, 1/2 and
1 ML coverages are discussed followed by the results for Te
adsorption on As adsorbed Si(211).  We then discuss the probable
replacement of surface Si atoms by As and finally the results for
the Te adsorption on the most favorable As-exposed Si(211) is
discussed. Note that one monolayer corresponds to one atom per
surface Si atom and the calculations presented here are obtained by
using the bulk terminated Si(211) as the starting structure prior to
As adsorption.

\subsection{Adsorption of As on Si(211)}
\label{subsec:AsAdsorption}

\subsubsection{1/8 ML As adsorption on Si(211)}

\begin{figure}
  \includegraphics[width=2.5in]{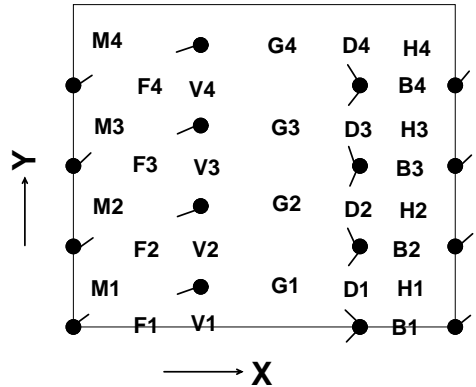}\\
  \caption{Filled circles represent Si atoms of the bulk terminated
  Si(211) surface. First, second, and third column of atoms
  are terrace, trench and edge atoms respectively. Atoms in the fourth
  column are the repetition of terrace atoms in the first column.
  Dangling bonds of all the atoms are shown. Various kinds of
  symmetric sites are designated as B, D, G, V, F, M and H sites
  respectively. The numerals 1, 2, 3, and 4 indicate identical
  sites displaced along the Y direction in the super-cell. For
  example, G1, G2, G3 and G4
  (G sites) are identical sites displaced by 3.84 \AA~ along the Y
  direction in the super-cell.}
  \label{fig3}
\end{figure}

\begin{figure}
  \includegraphics[width=2.5in]{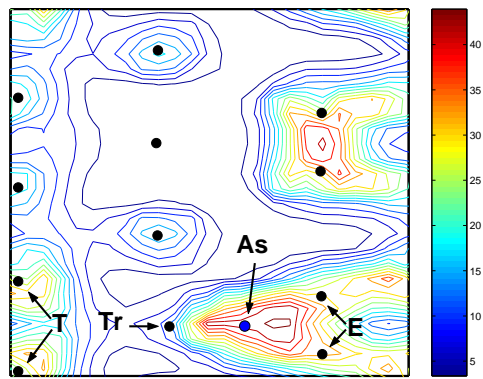}\\
  \caption{The total charge distribution on a plane just below the As
  layer for the most favorable configuration at 1/8 ML of As adsorption
  on Si(211). As position is indicated.}
  \label{fig4}
\end{figure}

Here we consider As adsorption at 1/8 ML coverage and hence, we need
to place one As atom on the surface of $1 \times 4$ super-cell. The
Si(211) surface offers various kinds of symmetric sites for As
adsorption. The different kinds of adsorption sites are shown in
Fig.~\ref{fig3} and they are designated as B, D, G, V, M, F and H
sites respectively. We will use numerals 1, 2, 3 and 4 to label
identical sites displaced along the $[01\bar{1}]$ direction in the
super-cell. For example, G1, G2, G3 and G4 (G sites) are identical
sites displaced by 3.84 \AA~ along the  $[01\bar{1}]$ direction in
the super-cell. The binding energy (BE) and height (from the top
layer of the Si surface) of the As atom at different kinds of sites
are given in Table \ref{table1}. The binding energy of the As atom
is defined as BE = E(As+Si) - E(Si) - ${\rm E_{As}}$ where E(As+Si),
E(Si) and ${\rm E_{As}}$ are total energy of As adsorbed super-cell,
total energy of the super-cell without As and atomic energy of As
respectively. From Table \ref{table1}, we notice that a G site
(BE=5.64 eV) is most favorable for As adsorption followed by a F
site (BE=5.43 eV), and a V site (BE=5.28 eV) respectively. Binding
energy of As at a V site is close to that at a F site while a B site
is less favorable. It is reasonable that the As atom favors to bind
at a G sites because it can satisfy itself by sharing charges with
three neighboring Si atoms (two edge Si and one trench Si atom) as
shown in Fig.~\ref{fig4}. Moreover, the distance of the As atom (at
a G site) from the nearest edge Si atoms are $\sim$ 2.45 \AA~ and
that from the nearest trench Si atom is $\sim$ 2.6 \AA. Though the
As atom at F or M sites can share its charge with three Si
neighbors, the As-Si bonds at those sites are not as strong as that
at a G site. For example, the distances of the As atom at a F site
from its neighboring Si atoms are 2.57 \AA~, 2.51 \AA~ and 2.62 \AA~
respectively which are larger compared to those for As at a G site.
We note that an attempt was made earlier to find the most favorable
site for As at 1/8 ML coverage. However, the G, F, and M sites were
not considered and therefore, their conclusion that a V site is most
favorable was based on the comparison among binding energies of As
atom at B, V, D, and H sites. Here we find that a V site is the
third favorable site for As. In our calculations, we let the As atom
(except at B, D and V sites) relax in all directions so as to reach
the local energy minima. For B and D sites, the As atom is allowed
to move only along the Z axis while that at a V site is allowed to
relax both along Y and Z directions. Here we conclude that at 1/8 ML
coverage, the As atom binds at a G site by sharing its charge with
two edge Si atoms and one trench Si atom. The charge density plot in
Fig.~\ref{fig4} clearly shows that the As atom at a G site (G1 site
is considered here) makes strong bond with the neighboring trench Si
atom along with the neighboring edge Si atoms by pulling the trench
atom by 0.4 \AA~ along the positive X direction. We also notice that
the edge Si atoms form dimers and thus reduces the total energy of
the system. A net energy gain in this process is the difference
between the binding energy of As and its chemical potential,
$\Delta$E = (BE - $\mu_{As}$) = 0.97 eV, where $\mu_{As} = 4.67$ eV
is the bulk chemical potential for Arsenic. In other words, if an As
atom is taken out of the As source and put at any of the G sites on
the Si(211) surface, an energy of 0.97 eV is gained.

\subsubsection{1/4 ML As adsorption on Si(211)}

\begin{figure}
  \includegraphics[width=2.5in]{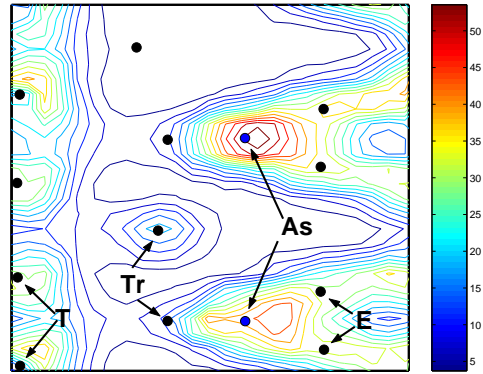}\\
  \caption{The total charge distribution just below the As layer for the most
  favorable configuration (G1G3) at 1/4 ML of As adsorption on Si(211). As positions
  are indicated.}
  \label{fig5}
\end{figure}

Here we consider As adsorption at 1/4 ML coverage, {\em i.e.}, two
As atoms are adsorbed on the bulk terminated Si(211) surface of the
super-cell. Based on the results at 1/8 ML coverage and on physical
grounds, we can chose all the reasonable combinations of a couple of
sites where As atoms may prefer to bind. The composite sites that we
consider are G1G3 (one As atom is placed at G1 site and the other is
placed at G3 site on the surface of the super-cell), B1B2 (one As
atom is adsorbed at B1 site and the other is adsorbed at B2 site),
G1F3 (one As atoms is adsorbed at G1 site and the other is adsorbed
at F3 site), G1M3 (one As atom is adsorbed at G1 site and the other
is adsorbed at M3 site), F1F3 (one atom is adsorbed at F1 site and
the other is adsorbed at F3 site) respectively. Without doing any
calculations and just based on the results at 1/8 ML coverage, one
may conclude that the composite site, G1G3 should be preferable for
As. However, in the presence of two As atoms the surface may undergo
further reconstructions due to the interaction with the surface Si
atoms and hence, G1G3 may not be the most favorable composite site.
We therefore, perform extensive calculations for As adsorption at
all the composite sites. The average binding energies per As atom
and the height of the As layer from the topmost Si layer are given
in Table \ref{table2}. By examining Table \ref{table2}, it turns out
that G1G3 configuration is indeed the most favorable one. The total
charge distribution on a plane just below the As layer for the G1G3
configuration is shown in Fig.~\ref{fig5}. We observe that the As
atoms at G1 and G3 sites form strong bonds with their neighbor Si
atoms. In addition, the edge Si atoms form dimers to reduce the
total energy of the system. The charge distribution for each As is
qualitatively similar to that shown in Fig.~\ref{fig4}. However, we
interestingly note that the average binding energy per As atom is
higher (by 0.18 eV) compared to that at 1/8 ML coverage. Generally,
the average binding energy per adsorbent atom increases when the
bond formation takes place among the adsorbent atoms. Here the
distance between two adsorbed As atoms is $\sim$ 7.68 \AA~ which
rules out any significant direct interaction between As atoms. Thus,
the increase of the average binding energy per As atom is due to an
overall reconstruction of the surface in terms of bond lengths and
charge sharing. From the symmetry of the system one would expect
that the alternative trench atoms should behave similarly. However,
from Fig.~\ref{fig5} we note that the alternative trench atoms
(trench Si atoms not bonding with the As atoms) do behave
differently; one of them (top most trench atom in Fig.~\ref{fig5})
is pushed down by $\sim$ 0.6 \AA~and also shifts towards negative X
direction by a small amount ($\sim$ 0.4 \AA).~ So, two As atoms
induce a reconstruction of the whole super-cell. From this
interesting result we conclude that the 1/4 ML coverage of As is
more favorable compared to 1/8 ML coverage. In other words, if the
Si(211) is exposed to As, the As coverage will readily approach 1/4
ML. The net energy gain per As atom in the process of bringing two
As atoms from the source and putting them at G1 and G3 sites on the
super-cell is ~ 1.15 eV.

\subsubsection{1/2 ML As adsorption on Si(211)}

\begin{figure}
  \includegraphics[width=2.5in]{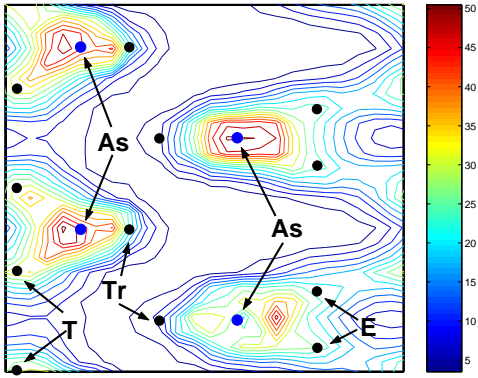}\\
  \caption{The charge distribution just below the As layer for the second
  favorable configuration (G1G3M2M4) at 1/2 ML of As adsorption on
  Si(211). As positions are indicated in the figure.}
  \label{fig6}
\end{figure}

\begin{figure}
  \includegraphics[width=2.5in]{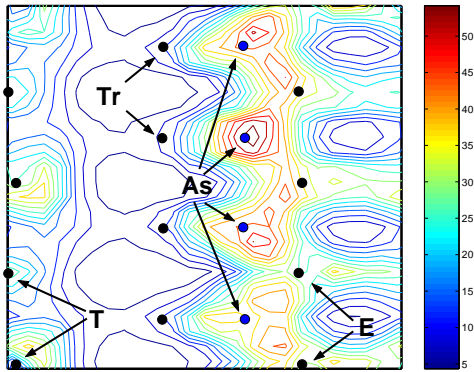}\\
  \caption{The charge distribution just below the As layer for the
  most favorable configuration (G1G2G3G4) at 1/2 ML of As
  adsorption on Si(211). As positions are indicated in the figure.}
  \label{fig7}
\end{figure}

Here we increase the As coverage to 1/2 ML. Based on previous
results at low coverages and the dangling bonds available on the
surface we consider three reasonable combinations of four sites on
the surface. They are designated as B1B2B3B4 (four As atoms are
adsorbed at B1, B2, B3 and B4 sites available on the surface of the
super-cell), G1G3M2M4 (four As atoms are adsorbed at G1, G3, M2 and
M4 sites available on the surface of the super-cell) and G1G2G3G4
(four As atoms are adsorbed at G1, G2, G3 and G4 sites available on
the surface of the super-cell) respectively. The average binding
energy per As atom and the height of the As plane from the top Si
layer for all the configurations are given in Table \ref{table3}. We
note that average energy gain per As atom for the configurations
G1G3M2M4 (BE $\sim$ 5.72 eV) and G1G2G3G4 (BE $\sim$ 5.80 eV) are
very close to each other. The charge density plot for the G1G3M2M4
configuration is shown in Fig.~\ref{fig6}. We note from
Fig.~\ref{fig6} that all the four As atoms make strong bonds with
their neighboring Si atoms. All the As atoms essentially remain on
the same plane and saturates all the surface dangling bonds.
Therefore, for the G1G3M2M4 configuration the highly terraced
Si(211) surface becomes flat decorated with parallel zigzag As
chains separated by 9.4 \AA~ and extended along the Y direction.
However, for the most favorable configuration, G1G2G3G4, the terrace
Si atoms remain unsaturated, all the As atoms remain on the same
plane and make strong bonds with the neighboring trench and edge Si
atoms. Thus, at 1/2 ML As coverage, the Si(211) surface becomes a
flat surface decorated with straight parallel As chains separated by
9.4 \AA~ and running along the Y direction (see charge density plot
in Fig.~\ref{fig7}). The average energy gain per As atom in the
process of bringing four As atom from the source and putting them at
G1,G2,G3, and G4 sites is ~1.14 eV which is substantial. The finding
of stable and straight As chains may be very useful in the context
of nanowire formation on the Si substrates. It is worth mentioning
that Ga adsorbed Si(211) surface has experimentally been observed
\cite{baski1} to be flat decorated with straight Ga chains extending
along the Y direction on Si(211). A comparison of the configurations
G1G3M2M4 and G1G2G3G4 also suggests that complete saturation of
surface dangling bonds does not always lead to the most favorable
structure. The least favorable configuration is B1B2B3B4. In this
configuration two As atoms at two nearby B sites form a bond. In
this B1B2B3B4 configuration, the As atoms reside on a plane at 2.1
\AA~ above the top Si layer and consequently, the surface roughness
increases to some extent and the dangling bonds of trench atoms
remain unsaturated. If As in dimer form is adsorbed, B1B2B3B4
configuration is the favorable one. Here we note that in an earlier
work \cite{sen} the B1B2B3B4 configuration was studied and the
average binding energy per As atom was found to be 5.28 eV which is
slightly less ($\sim$ 0.1 eV) compared to our result.

\subsubsection{1 ML As adsorption on Si(211)}

Here we consider 1ML coverage of As on the Si(211) surface. At 1 ML
coverage, we have limited number of choices for As configurations on
the Si(211). Based on physical grounds, {\em i.e.}, the symmetry of
the surface and the available dangling bonds on the Si(211) surface,
we consider two possible configurations on the bulk terminated
Si(211). The first configuration is represented as G1G2G3G4M1M2M3M4
where eight As atoms occupy four different G sites and four
different M sites available on the surface. In the second
configuration eight As atoms occupy just the next layer to the top
Si layer of the slab (the second Si layer from the bottom of the
slab is repeated on the top of the slab with As atoms). Our
calculations reveal that the second configuration is energetically
more stable than the first one. Average binding energy per As atom
in this case is 5.46 eV. Therefore, the energy gain per As atom in
the process of bringing eight As atoms from the source and placing
them on the Si(211) surface is $\sim$ 0.8 eV.

\begin{figure}
  \includegraphics[width=2.5in]{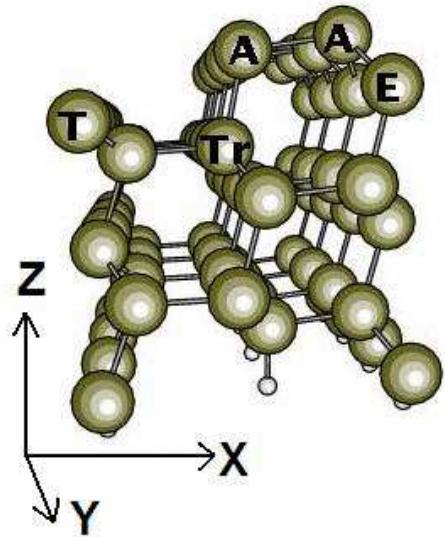}\\
  \caption{Atomic positions in the $1 \times 4$ super-cell
  after the adsorption of 1ML of As on Si(211). As atoms
  are denoted by A.}\label{fig8}
\end{figure}

The most favorable structure of the Si(211) with 1 ML of As
adsorption is shown in Fig.~\ref{fig8}. It is clear from the figure
that the bulk terminated Si(211) structure is retained with an added
layer of As (As atoms are denoted as A in Fig.~\ref{fig8}). For this
1 ML As adsorbed surface, the terrace atoms have one dangling bond
each and the As atoms nearer to the terrace atoms have one free
electron each while the dangling bonds of the edge atoms are
eliminated.

\subsection{Te on As adsorbed Si(211)}
\label{subsec:TeOnAsAdsorbedSi}

At this stage we have sound knowledge about the As adsorbed Si(211)
surface at various coverages. In MCT growth experiments, better
quality surfaces are obtained by adsorbing Te on the As-exposed
Si(211) for the essential growth of MCT. Thus, it is important to
investigate the atomic configuration of Te which forms the second
layer on the Si(211). As-exposed Si(211) surfaces that we consider
here are obtained by adsorbing As on Si(211). We consider only 1 ML
As adsorbed Si(211) surface because recent X-Ray photoemission
spectroscopy (XPS) data \cite{chad} indicated that the As coverage
on the Si(211) to be about 1 ML.

The 1 ML As adsorbed Si(211) surface (see Fig.~\ref{fig8}) offers
two distinct binding sites for Te. One is between two As atoms
sitting on the trench Si atoms (denote as S1 site) and the other one
is between the two terrace Si atoms (denote as S2 site). It is found
that the energy gain in bringing a single Te atom from source and
putting it at S1 is $\sim 2.7$ eV while that for S2 site is $\sim
2.1$ eV. Therefore, S1 site is more favorable at 1/8 ML coverage of
Te. When a second Te atom is placed at another S1 site available on
the surface in the presence of the first Te, the average energy gain
for each Te is $\sim 2.1$ eV. Similarly when third and fourth Te
atoms are placed at S2 sites available on the surface, the average
energy gains are $\sim 1.5$ eV and $\sim 1.1$ eV respectively. Here
we note that the energy gains are calculated by using the value of
-3.16 eV for the bulk chemical potential of Te. Thus one would
normally expect the Te coverage to be no less than 50\% of 1 ML.
However, in experiment, the Te coverage on the As-exposed Si(211)
was found to be 20\% - 30\%. \cite{nibir}. The discrepancy in the Te
coverage suggests that the As adsorbed Si(211) may not be the most
favorable surface. In other word, Te may not be adsorbing on the
As-exposed Si(211) substrate that we have considered so far.

\subsection{As Replaces Si on Si(211)}
\label{subsec:AsReplacement}

We have so far considered Te adsorption on As-exposed Si(211)
surfaces that were obtained by adsorbing As on the bulk terminated
Si(211). However, the Te coverage on such an As-exposed Si(211) with
1 ML of As does not agree with the experimental finding.
\cite{nibir} Borrowing from important results for Si(111) surface,
\cite{bringans} we investigate the possibility that Si atoms on the
Si(211) surface may be replaced by As atoms. \cite{brill} Here we
report the results of total energy calculations for the possible
replacement of Si(211) surface Si atoms by As. We consider an ideal
$1 \times 4$ Si(211) super-cell. As mentioned earlier, there are
three kinds of atoms on the Si(211) which are named as terrace,
trench and edge atoms. Replacement of one, two, four, and eight
surface Si atoms by As atoms correspond to 1/8, 1/4, 1/2 and 1 ML
coverage of As respectively. Separate calculations are carried out
for the replacement of one, two and four terrace Si atoms by As.
Similar calculations are done to replace trench and edge Si atoms by
As respectively. Finally four terrace Si atoms along with four
trench Si atoms are replaced to have 1 ML coverage of As on Si(211).
The results are given below.

The net energy gain in the process of replacing one terrace Si atom
by As is $\sim 1.1$ eV, one trench Si atom by As is $\sim 1.42$ eV
and an edge Si atom by As is $\sim 1.1$ eV respectively. We notice
that the replacement of a single trench Si atom by an As atom is the
most favorable while replacement of terrace and edge atoms are
equally less favorable. Comparing the energy gain due to the 1/8 ML
of As adsorption on Si(211) with the energy gains in the replacement
processes at 1/8 ML As coverage, we find that the replacement of any
surface Si atom by As is energetically more favorable relative to
the As adsorption.

The energy gain per As atom when two terrace Si atoms are replaced
by two As atoms, two trench Si atoms are replaced by two As atoms
and two edge Si atoms are replaced by two As atoms are 1 eV, 1.3 eV
and 0.95 eV respectively. Here the replacement of trench Si atoms
are most favorable while replacement of edge atoms are least
favorable. When comparing with the adsorption of As at 1/4 ML
coverage, we find that only the replacement of trench atoms are
energetically more favorable.

We next replace four terrace Si atoms by As, four Si trench atoms by
As and four edge Si atoms by As and corresponding energy gains per
As atom are 1.01 eV, 1.3 eV and 0.92 eV respectively. Here also we
find that the replacement of trench atoms are more favorable
compared to the adsorption of As at 1/2 ML coverage.

An interesting and useful case is the replacement of eight Si atoms
by eight As atoms which corresponds to 1 ML coverage of As because
recent experimental data indicates \cite{chad} that the As coverage
on Si(211) is about 1 ML. Based upon the previous results, we
replace four trench Si atoms along with four terrace Si atoms and we
find that the average energy gain per Si replacement is 1.17 eV
which is larger by 0.38 eV when compared with the adsorption of 1ML
As on Si(211). We therefore, conclude that the replacement of Si
atoms by As atoms are energetically more favorable compared to the
adsorption. In other words, irrespective of As coverage, the
As-exposed Si(211) surface obtained by replacing Si atoms with As is
more favorable compared to that obtained by adsorbing As on Si(211).
Thus, our results support the prevailing view \cite{brill} that the
Si atoms on Si(211) may be replaced by As atoms.

\subsection{Te on As replaced Si(211)}
\label{subsec:TeOnSiRepByAsSi}

It was mentioned earlier that Te is adsorbed on As-exposed Si(211)
in the MCT growth experiment. The As-exposed Si(211) surface
obtained by replacing all terrace and trench Si atoms of the bulk
terminated Si(211) by As (corresponding to 1 ML As coverage) is
considered here to study the Te adsorption. The As-exposed surface
will have sites like G1, G2, G3 and G4 as discussed earlier. Based
on our earlier results for the As adsorption on the bulk terminated
Si(211) and also available dangling bonds on the As-exposed Si(211)
surface, we conclude that a Te atom will prefer to bind at any of
the four G sites available on the surface. An energy of 1.32 eV is
gained when one Te atom is placed at G1 site. In presence of the
first Te atom, the second Te is placed at the G3 site available on
the surface of the super-cell (which corresponds to 1/4 ML coverage)
and an average energy of 1.34 eV is gained for each Te atom. At this
situation, two Te atoms at G1 and G3 sites make strong bonds with
neighboring edge Si atoms and the edge Si atoms form dimers to lower
the energy of the system. When the third Te is placed at G2 site in
the presence of two Te atoms at G1 and G3 sites (which corresponds
to 3/8 ML coverage) an average energy of $\sim$ 0.6 eV is gained for
each Te. We see that the average energy gain for each Te atom is
drastically reduced at 3/8 ML coverage of Te. This drastic drop in
the average energy gain for each Te indicates that a transition
takes place above 1/4 ML coverage. Though the average energy gain
per Te indicates a kind of transition at 3/8 ML coverage of Te, it
may not be give us a quantitative picture about the Te coverage on
the As-exposed Si(211) because G1, G2, and G3 sites on the surface
are no more identical in the presence of two Te atoms at G1 and G3
sites. We therefore, need to consider the energy gain for individual
Te atoms as long as all bonding sites are not identical. First Te
atom occupies the G1 site, bonds with two edge Si atoms and gain an
energy of $\sim$ 1.3 eV. In presence of the first Te, the second Te
occupies the G3 site, bonds with two other edge Si atoms and gains
an energy $\sim$ 1.3 eV. The individual energy gains for first and
second Te atoms are roughly same because Te atoms at G1 and G3 sites
have similar surroundings. Note that at 1/4 ML coverage of Te (G1
and G3 sites in the super-cell are occupied by two Te atoms)
dangling bonds of the edge Si atoms are completely saturated. This
implies that the third Te atom can not bind to the surface unless it
is capable of breaking the Si dimers that are formed at 1/4 ML
coverage of Te. In fact our calculation reveals that the third Te
atom placed at G2 site can not break the Si dimers. Furthermore, an
energy comparison shows that we need to supply $\sim$ 0.8 eV of
energy to bind the third Te atom on the As-exposed surface. This is
the reason behind a drastic drop in average energy gain for each Te
at 3/8 ML coverage of Te. Thus, 3/8 ML coverage of Te on the
As-exposed surface is highly improbable. However, when two Te atoms
are placed at G2 and G4 sites in presence of two Te atoms at G1 and
G3 sites (which corresponds to 1/2 ML coverage), the Si(211) the
edge Si dimers break to accommodate all four Te atoms. Under such
circumstances, all four Te atoms bond equally strongly with the edge
Si atoms. An average energy of 1.30 eV per Te atom is gained. Thus,
it appears that Te coverage between 1/4 ML and 1/2 ML is improbable
on the As-exposed Si(211). However, to have 1/2 ML coverage one has
to go through coverages between 1/4 and 1/2 ML. Therefore, Te
coverage beyond 1/4 ML (25 \% of ML) is improbable and this result
is in agreement with the experimental finding for the Te coverage
\cite{nibir} on the As-exposed Si(211).

\section{Summary}
\label{sec:summary}

Electronic structure calculations are performed to study the Te
coverage on the As-exposed Si(211) surfaces. The As adsorption on
Si(211) at 1/8, 1/4, 1/2 and 1 ML coverages are systematically
studied. Then the Te adsorption study on the 1 ML As adsorbed
Si(211) is done and it is found that 1/2 ML of Te adsorption is
energetically feasible on the one ML As adsorbed Si(211) surface.

We have done another set of calculations to verify the prevailing
idea that terrace and trench Si atoms on the Si(211) surface may be
replaced by As. We found that replacement of any surface Si atoms by
As atoms is energetically more favorable when compared with the
adsorption of As on the Si(211) at 1/8 ML coverage. At 1/4 and 1/2
ML coverages the replacement of trench Si atoms by As are more
favorable compared to the As adsorption on Si(211) while the
replacement of edge atoms are least favorable. The As-exposed
Si(211) with 1 ML of As is obtained by replacing all the trench and
terrace atoms by As and it is found to be more favorable compared to
that obtained by adsorbing 1 ML As on Si(211). The Te adsorption on
such a favorable As-exposed Si(211) is investigated. A drastic drop
in average energy gain per Te  is noted at 3/8 ML coverage of Te.
The drastic drop in the average energy gain suggests that the
probable Te coverage on the As-exposed Si(211) is 25\% of a ML which
is confirmed by further analysis.  This result agrees with the
experimental finding. \cite{nibir}

In conclusion, our results reveal that the replacement of surface Si
atoms on Si(211) by As is energetically more favorable when compared
with the As adsorption on Si(211). Thus our calculations support the
prevailing view that the terrace and trench Si atoms on the Si(211)
surface are replaced by As. The Te coverage on the As adsorbed
Si(211) may be more than 50\% of 1 ML which is way above the
experimentally found value. However, the Te coverage on the
energetically most favorable As-exposed Si(211)surface (surface
terrace and trench Si atoms replaced by As) is found to be 25\% of 1
ML and this is in agreement with the experimental finding.
\cite{nibir}

\acknowledgements{This work was supported under the research
agreement subcontract \#S03-16 from the Electro-Optics Center (EOC),
funded by the Marine Crops, monitored by Ray Balcerak and Kenneth
Freyvogel under the direction of Karl Harris. }

\newpage

\begin{table*}
\caption{Binding energies (BE) for As at various kinds of sites on
the bulk terminated Si(211) at 1/8 ML coverage. Here $\Delta$h is
the height of the adsorbed As atoms from the topmost Si layer on the
surface.} \label{table1}
\begin{ruledtabular}
\begin{tabular}{|c|c|c|c|c|c|c|c|}
  \hline
  {\bf Site:} & {\bf B1} & {\bf D1} & {\bf G1} & {\bf V1} & {\bf M1} & {\bf F1} & {\bf H1} \\
  \hline
  {\bf BE (eV):} & 4.66 & 4.35 & 5.64 & 5.28 & 5.12 & 5.43 & 4.04 \\
  \hline
  {\bf $\Delta$ h (\AA):} & 1.86 & 1.69  & 0.73 & 0.58 & 0.84 & 0.61 & 1.60 \\
  \hline
\end{tabular}
\end{ruledtabular}
\end{table*}

\begin{table*}
\caption{Average Binding energy (BE) per As atom at 1/4 ML coverage.
A set of two sites combinations on the bulk terminated Si(211) are
considered and they are denoted as B1B2, G1G3, G1F3, G1M3 and F1F3
respectively. Here $\Delta$h is the height of the adsorbed As atoms
from the topmost Si layer on the surface.} \label{table2}
\begin{ruledtabular}
\begin{tabular}{|c|c|c|c|c|c|}
  \hline
  {\bf Site:} & {\bf B1B2} & {\bf G1G3} & {\bf G1F3} & {\bf G1M3} & {\bf F1F3} \\
  \hline
  {\bf BE (eV):} & 5.34 & 5.82 & 5.66 & 5.11 & 5.00 \\
  \hline
  {\bf $\Delta$ h (\AA):} & 1.77 & 1.06 & 0.58 & 0.62 & 0.57  \\
  \hline
\end{tabular}
\end{ruledtabular}
\end{table*}

\begin{table*}
\caption{Average Binding energy (BE) per As atom at 1/2 ML coverage.
A set of four sites combinations on the bulk terminated Si(211) are
considered and they are denoted as B1B2B3B4, G1G3M2M4 and G1G2G3G4
respectively. Here $\Delta$h is the height of the adsorbed As layer
from the topmost Si layer on the surface.} \label{table3}
\begin{ruledtabular}
\begin{tabular}{|c|c|c|c|}
  \hline
  {\bf Site:} & {\bf B1B2B3B4} & {\bf G1G3M1M4} & {\bf G1G2G3G4} \\
  \hline
  {\bf BE (eV):} & 5.35 & 5.72 & 5.80 \\
  \hline
  {\bf $\Delta$ h (\AA):} & 2.10 & 0.96 & 0.32  \\
  \hline
\end{tabular}
\end{ruledtabular}
\end{table*}

\end{document}